\documentclass[12pt,preprint]{aastex}

\slugcomment{accepted for ApJ}

\shortauthors{Nishikawa et al.}

\shorttitle{Electron-Positron Relativistic Shocks}

\begin{document}

\title{Particle Acceleration and Magnetic Field Generation in Electron-Positron
  Relativistic Shocks}

\author{K.-I. Nishikawa\altaffilmark{1}}
\affil{National Space Science and Technology Center,
  Huntsville, AL 35805}
\email{ken-ichi.nishikawa@nsstc.nasa.gov}

\author{P. Hardee}
\affil{Department of Physics and Astronomy,
  The University of Alabama,
  Tuscaloosa, AL 35487}
\email{hardee@athena.astr.ua.edu}

\author{G. Richardson}
\affil{Department of Mechanical and Aerospace Engineering
University of Alabama in Huntsville
Huntsville, AL 35899}
\email{georgia.richardson@msfc.nasa.gov}

\author{R. Preece}
\affil{Department of Physics,
  University of Alabama in Huntsville,
  Huntsville, AL 35899 and National Space Science and Technology Center,
  Huntsville, AL 35805}

\author{H. Sol}
\affil{LUTH, Observatore de Paris-Meudon, 5 place Jules Jansen 92195
   Meudon Cedex, France}

\and

\author{G. J. Fishman}
\affil{NASA-Marshall Space Flight Center, \\
National Space Science and Technology Center,
  Huntsville, AL 35805}

\altaffiltext{1}{NRC Associate / NASA Marshall Space Flight Center \newpage
}


\begin{abstract}
\vspace{-0.5cm}

Shock acceleration is an ubiquitous phenomenon in astrophysical
plasmas.  Plasma waves and their associated instabilities (e.g.,
Buneman, Weibel and other two-stream instabilities) created in
collisionless shocks are responsible for particle (electron, positron,
and ion) acceleration. Using a 3-D relativistic electromagnetic
particle (REMP) code, we have investigated particle acceleration
associated with a relativistic electron-positron jet front propagating
into an ambient electron-positron plasma with and without initial
magnetic fields. We find small differences in the results for no
ambient and modest ambient magnetic fields. New simulations show that the
Weibel instability created in the collisionless shock front accelerates
jet and ambient particles both perpendicular and parallel to the jet
propagation direction.  Furthermore, the non-linear fluctuation amplitudes of
densities, currents, electric, and magnetic fields in the
electron-positron shock are larger than those found in the electron-ion
shock studied in a previous paper at the comparable simulation time. 
This comes from the fact that both
electrons and positrons contribute to generation of the Weibel
instability. Additionally, we have performed simulations with different
electron skin depths.  We find that
growth times scale inversely with the plasma frequency, and
the sizes of structures created by the Weibel
instability scale proportional to the electron skin depth.  This is the
expected result and indicates that the simulations have sufficient grid
resolution.  While some Fermi acceleration may occur at the jet front,
the majority of electron and positron acceleration takes place behind
the jet front and cannot be characterized as Fermi acceleration. The
simulation results show that the Weibel instability is responsible for
generating and amplifying nonuniform, small-scale magnetic fields
which contribute to the electron's (positron's) transverse deflection
behind the jet head.  This small scale magnetic field structure is
appropriate to the generation of ``jitter'' radiation from deflected
electrons (positrons) as opposed to synchrotron radiation.  The jitter
radiation has different properties than synchrotron radiation
calculated assuming a a uniform magnetic field. The jitter radiation
resulting from small scale magnetic field structures may be important
for understanding the complex time structure and spectral evolution
observed in gamma-ray bursts or other astrophysical sources containing
relativistic jets and relativistic collisionless shocks.

\end{abstract}

\vspace{-0.5cm}

\section{Introduction}

Nonthermal radiation observed from astrophysical systems containing
relativistic jets and shocks, e.g., active galactic nuclei (AGNs),
gamma-ray bursts (GRBs), and Galactic microquasar systems usually has
power-law emission spectra. In most of these systems, the emission is
thought to be generated by accelerated electrons through the
synchrotron and/or inverse Compton mechanisms. Radiation from these systems
is observed in the radio through the gamma-ray region.  Radiation in
optical and higher frequencies typically requires particle acceleration
in order to counter radiative losses.  It has been proposed that the
needed particle acceleration occurs in shocks produced by differences
in flow speed.

Fermi acceleration is the most widely known mechanism for the
acceleration of particles in astrophysical environments characterized
by a power-law spectrum. This mechanism for particle acceleration
relies on the shock jump conditions in relativistic shocks (e.g.,
Gallant 2002; Niemiec \& Oskowski 2004).
Most astrophysical shocks are collisionless since
dissipation is dominated by wave-particle interactions rather than
particle-particle collisions. Diffusive shock acceleration (DSA) relies
on repeated scattering of charged particles by magnetic irregularities
(Alfv\'en waves) to confine the particles near the shocks. However,
particle acceleration near relativistic shocks cannot be characterized
as DSA because the propagation of accelerated particles ahead of the
shock cannot be described by spatial diffusion. Anisotropies in the
angular distribution of the accelerated particles are large, and the
diffusion approximation for spatial transport does not apply
(Achterberg et al.\ 2001).

Previous microphysical analyses of the energy conversion in
relativistic pair outflows interacting with an interstellar medium
consisting of cold protons and electrons (e.g., Brainerd 2000;
Schlickeiser et al.\ 2002) have demonstrated that the beam excites both
electrostatic and low-frequency magnetohydrodynamic Alfv\'en-type waves
via a two-stream instability in the background plasma. This work has
also provided the time evolution of the distribution functions of beam
particles and the generated plasma wave turbulence power spectra. While
in these simulations the jet front showed some evidence of Fermi
acceleration, the main acceleration of electrons appeared to take place
in the downstream region (e.g., Brainerd 2000; Schlickeiser et
al.\ 2002; Ostrowski \& Bednarz 2002).  Further work in this area is
required if significant progress is to be made in unraveling
the important collisionless processes in relativistic shocks.

Particle-in-cell (PIC) simulations can shed light on the physical
mechanism of particle acceleration that occurs in the complicated
dynamics within relativistic shocks.  Recent PIC simulations using
injected relativistic electron-ion jets show that acceleration occurs
within the downstream jet, rather than by the scattering of particles
back and forth across the shock as in Fermi acceleration (Frederiksen
et al.\ 2003, 2004; Nishikawa et al.\ 2003), and Silva et al.\ (2003)
have presented simulations of the collision of two inter-penetrating
electron-positron plasma shells as a model of an astrophysical
collisionless shock. In the electron-positron simulations performed
with counter-streaming jets (Silva et al.\ 2003), shock dynamics
involving the propagating jet head (where Fermi acceleration may take
place) was not investigated.  In general, these independent simulations
have confirmed that relativistic jets excite the Weibel instability
(Weibel 1959).  The Weibel instability generates current filaments and
associated magnetic fields (Medvedev \& Loeb 1999; Brainerd 2000; Pruet
et al.\ 2001; Gruzinov 2001), and accelerates electrons (Silva et
al.\ 2003; Frederiksen et al.\ 2003, 2004; Nishikawa et al.\ 2003).

In this paper we present new simulation results of particle
acceleration and magnetic field generation for relativistic
electron-positron shocks using 3-D relativistic electromagnetic
particle-in-cell (REMP) simulations. These new simulation results are
compared to previous electron-ion results. In our new simulations, an
electron-positron relativistic jet with Lorentz factor, $\gamma = 5$
(corresponds to 2.5 MeV) is injected into an electron-positron plasma
in order to study the dynamics of a relativistic collisionless shock
both with and without an initial ambient magnetic field.
This particular choice of Lorentz factor is appropriate to the
production of an internal jet shock in AGN jets or GRB jets when the
high speed material has a Lorentz factor about ten times the Lorentz
factor of the low speed material.

In the collisionless shock generated behind the head of the
relativistic jet the Weibel instability is excited in the downstream
region.  The instability generates current filaments elongated along
the streaming direction and associated transverse magnetic fields.
Acceleration of electrons and positrons in the jet and ambient plasma
accompanies the development of the Weibel instability. In \S 2 the
simulation model and initial conditions are described. The simulation
results including comparisons with previous electron-ion simulations
(Nishikawa et al.\ 2003, hereafter paper I) are presented in \S 3, 
four cases are compared in \S 4, and in \S 5 we summarize and 
discuss the new results. 

\vspace{-0.5cm}
\section{Simulation Setup}

The code used in this study is a modified version of the TRISTAN code,
a relativistic electromagnetic particle (REMP) code (Buneman
1993).  Descriptions of PIC codes are presented in Dawson (1983),
Birdsall \& Langdon (1995), and Hickory \& Eastwood (1988). This code
has been used previously for many applications including astrophysical
plasmas (Zhao et al.\ 1994; Nishikawa et al.\ 1997a).

Three simulations were performed using an $85 \times 85 \times 160$
grid with a total of 55 to 85 million particles (27
particles$/$cell$/$species for the ambient plasma) and an electron skin
depth, $\lambda_{\rm ce} = c/\omega_{\rm pe} = 4.8\Delta$, where
$\omega_{\rm pe} = (4\pi e^{2}n_{\rm e}/m_{\rm e})^{1/2}$ is the
electron plasma frequency and $\Delta$ is the grid size. One simulation
was performed using an $85 \times 85 \times 320$ grid with a total of
180 million particles (27 particles$/$cell$/$species for the ambient
plasma) and an electron skin depth, $\lambda_{\rm ce} = c/\omega_{\rm
pe} = 9.6\Delta$.  In all simulations jets are injected at $z =
25\Delta$ in the positive $z$ direction. In all simulations radiating
boundary conditions were used on the planes at $z =0,z_{\rm max}$. Periodic
boundary conditions were used on all other boundaries (Buneman 1993).
The ambient and jet electron-positron plasma has mass ratio $m_{\rm
e}/m_{\rm p} \equiv m_{\rm e^-}/m_{\rm e^+} = 1$.  The electron thermal
velocity in the ambient plasma is $v_{\rm th} = 0.1c$ where $c$ is the
speed of light.

As in paper I, two kinds of jets have been simulated: a ``thin'' jet
with radius $r_{\rm jet} = 4\Delta$ and a ``flat" (thick) jet that
fills the computational domain in the transverse directions (infinite
width).   The thin jet is injected into a magnetized ambient plasma
with magnetic field parallel to the jet. In two flat jet simulations,
one is injected into an ambient plasma magnetized like the thin jet and
one is injected into an unmagnetized ambient plasma. In one additional
flat jet simulation on the longer grid a jet is injected into an unmagnetized
ambient plasma but with half the plasma frequency and twice the skin depth
so that $\lambda_{\rm ce} = c/\omega_{\rm pe} = 9.6\Delta$. The
choice of parameters and simulations allows comparison with previous
simulations (Silva et al.\ 2003; Frederiksen et al.\ 2003, 2004;
Nishikawa et al.\ 2003), and also provides an investigation of skin
depth, growth rate, and potential grid resolution effects.

\vspace{-0.5cm}
\section{Simulation results}

\subsection{Thin Jet Injection into Magnetized Ambient Plasma}

The electron number density of the thin jet is $2.98n_{\rm b}$, where
$n_{\rm b}$ is the density of ambient (background) electrons.  The
average jet velocity $v_{\rm j} = 0.9798c$, and the Lorentz factor is
5. The jet has thermal velocity $v_{\rm j, th} = 0.01c$.
The time step $t = 0.013/\omega_{\rm pe}$, the ratio
$\omega_{\rm pe}/\Omega_{\rm e} = 2.89$, and the Alfv\'en speed (for
electrons) $v_{\rm Ae} \equiv (\Omega_{\rm e}/\omega_{\rm pe})c =
0.346c$. Here $\Omega_{\rm e} = eB/\gamma m_{\rm e}$ is the electron 
cyclotron frequency.
With the speed of an Alfv\'en wave given by $v_A = [V_A^{2}/(1
+ V_{\rm A}^{2}/c^{2})]^{1/2}$ where $V_A \equiv
[B^{2}/4\pi (n_{\rm e} m_{\rm e} + n_{\rm p} m_{\rm p})]^{1/2} = 0.245c$,
the Alfv\'en Mach number $M_{\rm A} \equiv
v_{\rm j}/v_{\rm A} = 4.123$.  With a magnetosonic speed $v_{\rm ms} \equiv
(v_{\rm th}^{2} + v_{\rm A}^{2})^{1/2}$ the Magnetosonic Mach number $M_{\rm ms}
\equiv v_{\rm j}/v_{\rm ms} = 3.795$. At least approximately the
appropriate relativistic Mach numbers multiply these values by the
Lorentz factor.  Thus, in an MHD approximation we are dealing with a
high Mach number shock with $\gamma M >> 1$. The gyroradius of ambient
electrons and positrons with $v_{\perp} = v_{\rm th} = 0.1c$ is
$1.389\Delta = 0.289\lambda_{\rm ce}$.  All the basic parameters are the
same as in paper I except for the Alfv\'en wave speed in the ambient,
$v_{\rm A} \sim 0.075c$, which was reduced by the ion mass, $m_{\rm p} =
20m_{\rm e}$.

Figure 1 shows the jet electrons at simulation time $t =
22.1/\omega_{\rm pe}$.
%
%
The jet electrons are bunched along the jet direction and expanded
transversely due to a two-stream-type instability. In this simulation
the thin jet remains axisymmetric and behaves quite differently from
the twisted thin electron-ion jet simulated in paper I. Since the
radius of the thin jet is of the order of the electron skin depth and
underresolved, it is not clear if the Weibel instability is excited.
Nevertheless, the electron bunching seen in this case as opposed to the
electron twisting see in the electron-ion case illustrates the
potential effect of ion mass on the two-stream instabilities.  The
differences between the two thin jet cases suggest excitation of the
electrostatic two stream instability for the electron-positron jet and
ambient medium as opposed to the Buneman electron-ion drift instability
for the case of electron-ion jet and ambient medium.  Since the
diameter of the jet is too small compared to the skin depth and to
realistic jets, further study will be performed using flat jets that
fill the grid in the transverse direction.

\vspace{-0.5cm}
\subsection{Flat Jet Injection into Magnetized Ambient Plasma}

The electron number density of the flat jet is $0.741n_{\rm b}$.
Ambient parameters are the same as in the case of the thin jet. In this
case, the jet makes contact with the ambient plasma at a 2D interface
spanning the computational domain. Here only the dynamics of the
propagating jet head and shock region is studied. Effectively we study
a small uniform portion of a much larger shock. This simulation system
is different from simulations performed using counter-streaming equal
number density particles spanning the computational domain in the
transverse direction. The important differences between this type of
simulation and previous counter-streaming simulations is that the
evolution of the Weibel instability is examined in a more realistic
spatial way including the motion of the jet head, and we can have
different number densities in beam and ambient medium.

Electron density and current filaments resulting from development of
the Weibel instability behind the jet front are shown in Figure 2 at
time $t = 23.4/\omega_{\rm pe}$.
%
%
The electrons are deflected by the transverse magnetic fields ($B_{\rm
x}, B_{\rm y}$) via the Lorentz force: $-e({\bf v} \times {\bf B})$,
generated by current filaments ($J_{\rm z}$), which in turn enhance the
transverse magnetic fields (Weibel 1959; Medvedev and Loeb 1999). The
complicated filamented structures resulting from the Weibel instability
have diameters on the order of the electron skin depth ($\lambda_{\rm ce} =
4.8\Delta$). This is in good agreement with the prediction of $\lambda
\approx 2^{1/4}c\gamma_{\rm th}^{1/2}/\omega_{\rm pe} \approx
1.188\lambda_{\rm ce} = 5.7\Delta$ (Medvedev \& Loeb 1999). Here,
$\gamma_{\rm th} \sim 1$ is a thermal Lorentz factor. The filaments are
elongated along the direction of the jet (the $z$-direction, horizontal
in Figure 2). The magnetic field and transverse current ($J_{\rm x}$) shows
significantly more transverse variation than was seen for the comparable
electron-ion jet (Figure 2 in paper I).

The acceleration of electrons has been reported in previous work
(Silva et al.\ 2003; Frederiksen et al.\ 2003, 2004; Nishikawa et
al.\ 2003; Hededal et al. 2004) and is shown here in Figure 3.
%
%
We see that the kinetic energy (parallel velocity $v_{\parallel}
\approx v_{\rm j}$) of the jet electrons is transferred to the
perpendicular velocity via the electric and magnetic fields
generated by the Weibel instability. The strongest transverse
acceleration of jet electrons (Fig.\ 3a) accompanies the strongest
deceleration of electron flow (Fig.\ 3b) and occurs between
$z/\Delta = 100 - 120$.  The transverse acceleration seen here is
over four times that seen in the comparable electron-ion
simulation in paper I at comparable simulation time
(see Fig.\ 3 paper I) and the deceleration
is also much greater. The strongest acceleration and deceleration
takes place around the maximum amplitude of perturbations due to
the Weibel instability at $z/\Delta \sim 105$ revealed
qualitatively in Figure 2 and more quantitatively in Figure 4.
Since the electrons and positrons have the same mass, they are
accelerated equally perpendicular to the ambient magnetic field.
At the jet front some jet electrons and positrons are accelerated
and some are decelerated.  This acceleration and deceleration is
indicated by the slanting of the parallel velocity distribution at
the jet head (Fig.\ 3b  at $z = 136\Delta$). Furthermore, slight
acceleration is found just behind the jet front ($z/\Delta \sim
130$). The positrons also have similar distributions (not shown).
However, in paper I for the electron-ion case, only electrons have
similar distributions as shown in Fig. 3b as ions have not had time to
react. This fact is consistent
with the electric field generated just behind the jet front.
This may indicate that some Fermi acceleration is taking place at the
jet front as described in previous work (e.g., Achterberg et al.\ 2001;
Gallant 2002; Ellison \& Double 2002), however, further investigation
is necessary. Figure 3 suggests that the ``acceleration region" has 
a thickness in the range $z/\Delta = 70 - 130$ behind the
front defined by the fastest moving jet electrons.  Possibly, the
``turbulence" assumed for the diffusive shock acceleration (DSA)
corresponds to this shock region (downstream but not upstream).

The ambient electrons and positrons are also accelerated, e.g.,
Fig.\ 3c \& 3d.  Some of the ambient electrons are accelerated
perpendicularly up to $0.6c$ and are accelerated in the direction of
jet flow to greater than $0.6c$.  The leading edge as defined by the
fastest jet electrons is not significantly reduced as would be the case
for the jet head in an ideal relativistic MHD simulation (e.g.,
Nishikawa et al.\ 1997b). An ideal MHD simulation would give a head
advance speed
$$
v_{\rm h} \sim {\gamma \eta^{1/2} \over 1 + \gamma \eta^{1/2}}v_{j} = 0.81c~,
$$
where $\eta = n_{\rm j}/n_{\rm b} = 0.741$ and we have ignored magnetic
and thermal pressures in the ram pressure balance equation (Mart\'{\i}
et al.\ 1997; Rosen et al. 1999).  On the other hand, the average
forward motion of the most decelerated jet particles and most
accelerated ambient particles is of this order.

Figure 4 shows 1-D cuts through the computational grid parallel to the
$z$-axis at $x/\Delta =$~ 38 and three locations $y/\Delta =$~ 38, 43,
and 48 separated approximately by the electron skin depth
($\lambda_{\rm ce} \sim 4.8\Delta$).
%
%
This figure provides some quantitative longitudinal information about
the filament structures shown qualitatively in Figure 2.  With
separation by about a skin depth the phase of the instability is
different along different cuts, but the amplitudes are similar.  The
growth time of the Weibel instability is calculated to be, $\tau
\approx \gamma_{\rm sh}^{1/2}/\omega_{\rm pe}$ (Medvedev \& Loeb 1999)
and here $\tau \approx 2.2/\omega_{\rm pe}$ with $\gamma_{\rm sh} = 5$.
If this is converted into a growth length $\ell \equiv c\tau = 2.2
\lambda_{\rm ce} \sim 10.7 \Delta$.  The simulation results show that
the maximum amplitudes are achieved at $z \sim 105 \Delta$ about
$80\Delta$ from the position of the jet injection at
$z = 25\Delta$. This result indicates that the Weibel instability
grows to maximum amplitude from thermal fluctuations in about eight
growth lengths (eight growth times) at $t = 23.4/\omega_{\rm pe}$.
The electron density shown in Fig.\ 4a indicates that the
width of the jet head is slightly larger than the electron skin depth,
$4.8\Delta$.  A similar feature is not obvious in the electron-ion case
(see Fig.\ 4a in paper I).  The fluctuation amplitudes in the nonlinear
stage for the electron-positron case shown here in Figure 4 are much
larger than in the electron-ion case considered in paper I at the same time
(see Fig.\ 4
in paper I).  The electron density fluctuates by nearly a factor 2
about the average, whereas in the electron-ion case the fluctuation was
by less than a factor 1.2 about the average.

The $z$-component of current density shown in Fig.\ 4b indicates both
positive and negative currents in the jet head region and shows no
evidence for  the small negative current found at the leading edge of
the electron-ion jet in paper I.  In the electron-ion ``shock'' some
jet electrons are ahead of the ions.  Fluctuations in this component of
the current density are up to a factor 3 times larger than in the
electron-ion case.
Here the electric field amplitude is up to 4 times greater than that
found in the electron-ion case.  The induced transverse magnetic fields
are up to 10 times those found in the electron-ion case.  Based on
Figs.\ 2 and 4 the length of filaments along the jet, $\sim 10\Delta$,
around $z = 100\Delta$ is approximately twice the electron skin depth.
This result is consistent with the previous electron-positron
simulations performed by Silva et al.\ (2003).

Transverse structure accompanying the Weibel instability
is shown by 2-D slices of the longitudinal current density and
electron density along with the transverse magnetic field
in Figure 5.  
%
%
The size of these structures transverse to the jet
propagation is about the electron skin depth. Such transverse
structures are also found in counter-streaming jet simulations (Silva
et al.\ 2003; Frederiksen et al.\ 2003, 2004).  
The simulation results also show that smaller scale filaments
have merged into larger scale filaments in the nonlinear stage at the maximum
amplitudes (see also Silva et al.\ 2003; Frederiksen et al. 2004; 
Medvedev et al. 2004).

\vspace{-0.5cm}
\subsection{Unmagnetized Ambient Plasma and Electron Skin Depth}

The simulations of Silva et al.\ (2003) and Frederiksen et al.\ (2003,
2004) did not include ambient magnetic fields. We have performed one
flat jet simulation without ambient magnetic fields but otherwise identical to
the flat electron-positron jet injection into a magnetized
electron-positron plasma.  Here we can compare results with these
previous simulations, with our electron-ion jet injection into an
unmagnetized electron-ion plasma in paper I and evaluate the effect of
ambient magnetic fields on the perturbations. Additionally, we have
performed one flat jet simulation using a different electron skin depth,
$\lambda_{\rm ce} = c/\omega_{\rm pe} = 9.6\Delta$.   Here the electron
plasma frequency is half of that used in the original simulations, and
the system size is two times longer than the original size.  This
allows us to evaluate the effect of the electron skin depth on the size
of structures, the dependence of growth rate on the plasma frequency,
and the potential effect of our grid scale on the results.

The filamentary structure of electron density, magnetic fields and
currents resulting from development of the Weibel instability is shown
in Figure 6a \& 6b for $\lambda_{\rm ce} = 4.8\Delta$ and in Figure 6c
\& 6d for $\lambda_{\rm ce} = 9.6\Delta$.
%
%
Comparison between Figs.\ 6a \& 6b and Figure 2 reveals little
qualitative change in density, current or magnetic structure resulting
from an ambient magnetic field.   Quantitatively, the peak values of
the perturbations due to the Weibel instability for an unmagnetized
ambient plasma are somewhat larger, 20 - 25\%, than those for a
magnetized ambient plasma.  Comparison of Figs.\ 6a \& 6b with
Figs.\ 6c \& 6d shows that a doubling of the electron skin depth has
resulted in a predicted doubling of the size of structures, both
transversely and longitudinally.  Filamentary structures appear about
twice as far behind the leading edge defined by the fastest moving jet
particles. This doubling of structure size is the expected result if
structures scale with the electron skin depth
(Medvedev \& Loeb 1999).  The fact
that we find this scaling indicates sufficient grid resolution when
$\lambda_{\rm ce} = 4.8\Delta$ and with $\lambda_{\rm Debye} \sim
0.5\Delta$.

Transverse structure in the electron and current density, and in the
transverse magnetic field for the two different electron skin depths
is shown in a 2-D slice in the $x - y$
plane at $t = 23.4/\omega_{\rm pe}$ in Figure 7. Figures 7a and 7b 
correspond to Figs. 5a and 5b, which show that the weak initial 
ambient magnetic field does not affect on the evolution of Weibel
instability.
%
%
The grid size is the same for all panels, it is easy to see that the
structures with twice the electron skin depth are approximately two
times larger. The increase in size makes it easy to see (Fig.\ 7d) that
the transverse magnetic field ($B_{\rm x,\, y}$) is toroidal around the
current filaments ($J_{\rm z}$) represented by the color contours.
In internal shocks,
typical plasma densities of $n_{\rm e} \sim 3 \times 10^{10}~{\rm cm}^{-3}$, 
shock Lorentz factors $\gamma_{\rm sh} \sim 4$ and initial thermal Lorentz
factors, $\gamma_{\rm th} \sim 2$, yield plasma frequencies $\omega_{\rm e}
\sim 1 \times 10^{9}$, and electron skin depth $\sim 50$~cm  (eq. 9a in
Medvedev \&
Loeb (1999)).  At least approximately this gives an indication as to our
filament size in the shock reference frame in a GRB.  This scale is
much smaller than any observed spatial scale associated with the
source. In an external shock $n_{\rm e} \approx 4.3 {\rm cm}^{-3}, \gamma
_{\rm sh} \approx 39$, the relativistic electron skin depth 
($= c \gamma_{\rm sh}^{1/2}/\omega_{\rm e}$) is 
$\sim 10^{6}$~cm (Frail et al. 2004).

The longitudinal structure of perturbations along the $z$-direction in
$J_{\rm z}$, and $B_{\rm x}$ for the two different skin depths is
shown in Figure 8.  Here the 1-D cuts through the computational grid
parallel to the $z$-axis are located at $y/\Delta =$~ 38, 43, and 48
($\lambda_{\rm ce} = 4.8\Delta$) or  $y/\Delta =$~ 33, 43, and 53
($\lambda_{\rm ce} = 9.6\Delta$) and are separated approximately by an
electron skin depth.
%
%
Here we see that the lengths of filamentary structures are doubled by
the doubled skin depth as expected.  The only obvious difference
is a reduction in the maximum transverse magnetic field by almost a
factor of two.  Note also some accompanying reduction in the typical
current density, $J_{\rm z}$.  Since current density maxima are
comparable for both skin depths it seems likely that differences here
are largely an accidental result of the location of the cuts. Note how
similar the results seen in Figs.\ 8a and 8b are to those shown in
Figs.\ 4b \& 4c with an initial ambient magnetic field.  The modest
differences are a result of the different seed perturbations (thermal
noise caused by the initial loading of particles).

\vspace{-0.5cm}
\section{Electron-Positron, Electron-Ion Results Compared}

The efficiencies of conversion of bulk kinetic energy into radiation via
synchrotron or ``jitter" emission from relativistic shocks will be
determined by the magnetic field strength and the electron energy
distribution behind the shock.  In what follows we examine the
conversion of bulk kinetic energy into magnetic and thermal energy by
comparing the relevant energy densities in a volume consisting of
a number of cells. The simulations show that the initial jet bulk
kinetic energy is converted into magnetic energy, transverse
acceleration of the jet and ambient particles (thermal energy), and
acceleration of the ambient plasma through the Weibel instability. 

In order to compare characteristics of Weibel instabilities, 
we have evaluated the magnetic field energy, ambient electron
thermal energy and jet electron thermal energy for
four different cases at $t = 23.4/\omega_{\rm pe}$, all without initial
ambient magnetic fields. For the electron-positron jet and ambient we
calculated these quantities for the two different skin depths (case A
(smaller skin depth) and case B (larger skin depth)) considered in this
paper as shown in Figures 6, 7, and 8.  The values obtained here are
compared to similar values obtained for electron-ion jet and ambient
considered in paper I (case C) shown in Figures 7, 8, and 9 in paper
I.  Additionally, we have performed a new electron-ion simulation with
a larger skin depth (case D) that can be compared to case B in this
paper.  The volumes over which averages are determined include all
cells to the transverse boundaries of the grid between limiting $z$
distances.  For the larger skin depth (cases B and D) shocked
quantities are calculated in the region $155 < z/\Delta < 215$ where
the Weibel instability has largest amplitudes, and ambient quantities
are calculated in the injection region $25 < z/\Delta < 85$ where the
Weibel instability has not been excited. For cases A and C with the
smaller skin depth, shocked quantities are calculated in the region $90
< z/\Delta < 120$ and ambient quantities are obtained in the region $25
< z/\Delta < 55$.

The thermal energy contained in the chosen volume is given by
$U_{\rm th} = \sum (\gamma_{\rm th} - 1) m_{\rm n} c^{2}$ where
$\gamma_{\rm th} = [1 - (v_{\rm th}/c)^{2}]^{-1/2}$ and  where the
summation is over the number of particles, $m_{\rm n}$ is the mass
of particle $n$, and $v_{\rm th} = [(v^{\rm th}_{\parallel})^{2}
+(v^{\rm th}_{\perp})^{2}]^{1/2}$, is the thermal velocity for
particle $n$. Here we define the components of the individual
particle's velocity parallel to the bulk velocity,
$v_{\parallel} \equiv ({\bf v}\cdot
{\bf V})/V$, and perpendicular to the bulk velocity, $v_{\perp} \equiv
\mid\!{\bf v}\times {\bf V}\mid\!/V$, where ${\bf v}$ and ${\bf V}$
represent the motion of particle $n$ and the bulk motion of
particles, respectively ($V = [V^{2}_x + V^{2}_y + V^{2}_z]^{1/2}$).
We define the velocity
of the bulk motion ${\bf V} = \sum {\bf v}/n_{\rm cell}$,
where we sum the velocities, {\bf v}, over $n_{\rm cell}$ number of particles
in the grid zone.
The parallel and perpendicular components of the thermal velocity are
given by
$$
{v}^{\rm th}_{\parallel} = {v_{\parallel} - {V} \over 1 - {\bf
v}\cdot {\bf V}/c^{2}}~,
$$
and
$$
{v}^{\rm th}_{\perp} = {v_{\perp} \over \Gamma_{\rm V} (1 - {\bf
v}\cdot {\bf V}/c^{2})}~,
$$
where $\Gamma_{\rm V} = [1 - ({V}/c)^{2}]^{-1/2}$.
In general, we must separately compute bulk and thermal energy
for each particle species as bulk motion and thermal
velocity can be different for each species, here electrons, $e$,
positrons, $p$, and ions, $i$.  For our purposes it can also be
useful to separate the ambient, $a$, particles from the jet, $j$,
particles and to compare initial, $in$, and shocked, $sh$, states.
For example, $U^{\rm sh}_{\rm B} = {{B}_{\rm sh}}^{2}/8\pi$,
represents the shocked value of the magnetic energy, and $U^{\rm
e,j,sh}_{\rm th} = \sum (\gamma_{\rm th} - 1) m_{\rm e} c^{2} $,
where $\gamma_{\rm th} = [1 - ({v}^{\rm e,j,sh}_{\rm
th}/c)^{2}]^{-1/2}$ and $U^{\rm e,j,sh}_{\rm V} = \sum
(\Gamma_{\rm V} - 1) m_{\rm e} c^{2}$, where $\Gamma_{\rm V} = [1
- ({V}^{\rm e,j,sh}/c)^{2}]^{-1/2}$, represent the shocked values
of the jet electron thermal energy, and jet electron bulk kinetic
energy, respectively. The total kinetic energy is written $U^{\rm
e,j,sh}_{\rm k} = \sum (\gamma_{\rm k} - 1) m_{\rm e} c^{2} $,
where $\gamma_{\rm k} = [1 - ({v}^{\rm e,j,sh}/c)^{2}]^{-1/2}$.

At the comparable simulation time
perturbations associated with the Weibel instability grow to a larger
amplitude for the electron-positron jet and ambient than for the
comparable electron-ion case. A comparison between the magnetic field
energy, $U^{sh}_{\rm B}$, in the shock region (all values in
simulation units) for electron-positron, and electron-ion cases shows:
(A) $U^{\rm sh}_{\rm B} = 4.484 \times 10^4$,
(B) $U^{\rm sh}_{\rm B} = 4.088 \times 10^4$,
(C) $U^{\rm sh}_{\rm B} = 2.884 \times 10^3$, and
(D) $U^{\rm sh}_{\rm B} = 1.824 \times 10^3$.
The values in the cases A and C are multiplied by two since the volume
over which these numbers are obtained is half of that used for cases B
and D. This comparison reveals that the magnetic energy growth
accompanying growth of the Weibel instability is about 20 times larger
for electron-positron jet and ambient than for the comparable
electron-ion cases.

The increase in magnetic field energy above the initial thermal
fluctuations is calculated to be:
(A) $U^{\rm sh}_{\rm B}/U^{\rm in}_{\rm B} = 4.860 \times 10^3$,
(B) $U^{\rm sh}_{\rm B}/U^{\rm in}_{\rm B} = 6.080 \times 10^3$,
(C) $U^{\rm sh}_{\rm B}/U^{\rm in}_{\rm B} = 1.646 \times 10^3$, and
(D) $U^{\rm sh}_{\rm B}/U^{\rm in}_{\rm B} = 1.140 \times 10^3$.
These values show that the magnetic field energy for
electron-positron jet and ambient plasma is increased about four times more
than for the comparable electron-ion ($m_{\rm i}/m_{\rm e} = 20$) jet
and ambient plasma.

The increase in thermal energy of the jet electrons for the four
cases is calculated using the above definitions to be:
(A) $U^{\rm e, j, sh}_{\rm th}/U^{\rm e, j, in}_{\rm th} = 16.13$,
(B) $U^{\rm e, j, sh}_{\rm th}/U^{\rm e, j, in}_{\rm th} = 12.81$,
(C) $U^{\rm e, j, sh}_{\rm th}/U^{\rm e, j, in}_{\rm th} = 1.73$, and
(D) $U^{\rm e, j, sh}_{\rm th}/U^{\rm e, j, in}_{\rm th} = 1.31$.
The thermal energy of jet electrons in the electron-positron
jet is increased about ten times more than in the comparable
electron-ion jet.

If we compare the magnetic field energy in the shocked region
to the initial total kinetic energy , i.e., $\epsilon^{\rm
k}_{\rm B} = U^{\rm sh}_{\rm B}/(U^{\rm j, in} _{\rm k} + U^{\rm a,
in}_{\rm k})$, we find:
(A) $\epsilon^{\rm k}_{\rm B} = 1.01 \times 10^{-2}$,
(B) $\epsilon^{\rm k}_{\rm B} = 1.02 \times 10^{-2}$,
(C) $\epsilon^{\rm k}_{\rm B} = 0.72 \times 10^{-4}$, and
(D) $\epsilon^{\rm k}_{\rm B} = 0.45 \times 10^{-4}$.
For the electron-ion jet and ambient plasma $\epsilon^{\rm k}_{\rm B}
\approx 10^{-4}$ in agreement with the linear theory
predictions made by Wiersma \& Achterberg (2004). However, it should
be noted that in our electron-ion
simulations the mass ratio $m_{\rm i}/m_{\rm e} = 20$ and nonlinear
saturation has not been fully achieved and this may explain the
agreement with a linear prediction.
In any event, for the
electron-positron jet and ambient plasma $\epsilon^{\rm k}_{\rm B}
\approx 10^{-2}$, which may be in the nonlinear phase and is at least
two orders of magnitude larger. This
result is consistent with the efficiencies required by the
observed synchrotron radiation (Gruzinov \& Waxman 1999).

For the case B the average jet electron velocity ($V_{\rm z}$) in the
shock region ($155 < z/\Delta < 215$) is reduced to $V_{\rm z} = 0.9656 c$
from an initial $V_{\rm z} = 0.9765 c$
($25 < z/\Delta < 85$).  This result indicates the slow down
accompanying excitation of the Weibel instability.  Calculation of the
decrease in jet electron bulk kinetic energy, $\epsilon_{\rm V}
\equiv U^{\rm e,j,sh}_{\rm V}/U^{\rm e,j,in}_{\rm V}$, for the four
cases reveals:
(A) $\epsilon_{\rm V} = 0.737$,
(B) $\epsilon_{\rm V} = 0.805$,
(C) $\epsilon_{\rm V} = 0.976$, and
(D) $\epsilon_{\rm V} = 0.990$.
We see at least a 20 - 25\% conversion of bulk kinetic jet energy into
magnetic fields, thermalization of jet and ambient plasma, and
acceleration of ambient plasma for an electron-positron jet and
ambient.
The maximum efficiency for the electron-ion jet  at this simulation time
is 1 - 3\%.  During these simulation times
ions do not slow down with the electrons  and the
efficiency is reduced by a factor up to $m_e/m_i = 20$ as the total
initial bulk kinetic energy of electrons and ions is about 20
times larger. In order to fully take account of ion involvement in the
Weibel instability, simulations with a longer system and a longer
simulation time are required ($(m_{\rm i}/m_{\rm e})^{1/2}
\sim 4 - 5$) (Frederiksen et al. 2004; Hededal et al. 2004).

Calculation of the increase in thermal energy of ambient electrons
for the four cases shows that:
(A) $U^{\rm e, a, sh}_{\rm th}/U^{e, a, in}_{\rm th} = 8.38$,
(B) $U^{\rm e, a, sh}_{\rm th}/U^{e, a, in}_{\rm th} = 8.47$,
(C) $U^{\rm e, a, sh}_{\rm th}/U^{e, a, in}_{\rm th} = 1.32$, and
(D) $U^{\rm e, a, sh}_{\rm th}/U^{e, a, in}_{\rm th} = 1.22$.
In the electron-positron ambient plasma, the thermal energy of
electrons and positrons is increased about 6 times more than thermal
energy of the electrons in the electron-ion ambient plasma.
The electron-positron shock is much quicker
at converting bulk kinetic energy into thermal energy
(particle acceleration) and magnetic field energy.

\vspace{-0.5cm}
\section{Summary and Discussion}

We have performed self-consistent, three-dimensional relativistic
particle simulations of relativistic electron-positron jets propagating
into magnetized and unmagnetized electron-positron ambient plasmas.
The main acceleration of electrons takes place in the downstream region.
Processes in the relativistic collisionless shock are dominated by
structures produced by the Weibel instability.  This instability is
excited in the downstream region behind the jet head, where electron
density perturbations lead to the formation of current filaments. The
nonuniform electric field and magnetic field structures associated with
these current filaments decelerate the jet electrons and positrons,
while accelerating the ambient electrons and positrons, and
accelerating (heating) the jet and ambient electrons and positrons in
the transverse direction.

Two new findings are confirmed in this study. As shown in the previous sections,
density, current, electric and magnetic field amplitudes for electron-positron 
jet and plasma are significantly larger than the those for a comparable 
electron-ion jet and plasma at similar simulation time. In
the electron-positron plasma, both electron and positrons participate
in exciting the Weibel instability.  Transverse acceleration and
deceleration of the jet particles, an acceleration of the ambient
particles, and transverse acceleration (heating) of jet and ambient
particles inside the ``shock'' region is significantly larger than in
the electron-ion case as described in the previous sections. However, 
we evaluate our electron-ion results at the similar simulation time to our 
electron-positron results. At this simulation time ions have not yet 
participated in the dynamics significantly. At long time and longer 
spatial scales, the ion dynamics becomes dominant (e.g., Hededal et al. 
2004).

Secondly, comparison between simulations with different plasma frequency
reveals the expected growth rate decrease as the plasma frequency
decreases. This is accompanied by the expected growth length and
filament size increase as the electron skin depth, $\lambda_{\rm ce}
\propto \omega_{\rm pe}^{-1}$, increases as shown in Figs, 6, 7, and 8.

The Weibel instability originates from the fact that the electrons are
deflected by the perturbed (small) transverse magnetic fields
accompanying the current filaments and subsequently enhance the current
filaments  (Weibel 1959; Medvedev \& Loeb 1999; Brainerd 2000; Gruzinov
2001). The deflection of particles due to the Lorentz force increases
as the magnetic field perturbation grows in amplitude. Our results here
are consistent with results from previous simulations (Silva et
al.\ 2003; Frederiksen et al.\ 2003, 2004).

The basic nature of the Weibel instability (Medvedev \& Loeb 1999; 
Nishikawa et al.\ 2003) is also confirmed in this study of electron-positron
jet cases. In particular, the aperiodic nature of the instability 
($\omega_{\rm real}$ = 0 (convective)) is observed in the evolution 
of the generated transverse magnetic field ($B_{\rm y}$).
Thus, it can be saturated only by nonlinear effects and not by kinetic effects,
such as collisionless damping or resonance broadening. Hence the magnetic
field can be amplified to very high values locally as shown in Figs. 7b and 7d..

In general, we find that the absence of an ambient magnetic field leads
to slightly larger maximum values for the perturbations produced by the
Weibel instability.  Qualitatively there appears to be little change in
the current filament and transverse magnetic field structure resulting
from the ambient magnetic field that we have considered here. This
result is similar to that found previously for the flat electron-ion
jet injected into a magnetized and unmagnetized electron-ion plasma.
Thus, our present ambient magnetic field oriented parallel to the flow
direction with Alfv\'en wave speed greater than the thermal speed for
the the electron-positron plasma or less than the thermal speed for the
electron-ion plasma, has only a minor influence on the results.

The perturbed electron density and filamented currents have a
complicated three-dimen-sional structure. The transverse size of these
structures is on the order of the electron skin depth and is somewhat
larger if there are no ambient magnetic fields. However, the length of
structures along the jet direction is slightly larger than the
transverse scale.  At the termination of our simulations, for an
electron skin depth $\lambda_{\rm ce}=4.8\Delta$ the thickness of the
unstable region along the jet direction ranges from $z/\Delta = 55$ to
$135$ , is $\gtrsim 15 \lambda_{\rm ce}$, and is similar for electron-ion jets
and plasma.

The perturbation size in the transverse direction become largest around
$z/\Delta = 105$ where nonlinear effects lead to the merging of the
smaller scale filaments that first appear behind the jet front. This
result is similar to previous counter-streaming simulations (Silva et
al. 2003; Frederiksen et al. 2004; Medvedev et al. 2004) in which 
smaller filaments first appear and then merge into
larger filaments at a later time. Now we see the temporal development
appear as a spatial development and this occurs about eight growth
lengths, $8 \ell \sim 6.5c/\omega_{\rm pe} \sim 80\Delta$ from the
position of the jet injection.

Recent observations show that from optical observations alone the
wiggles in the light curves of GRB 011211 are the result of spherically
asymmetric density or energy variations, i.e. variations that cover
less than the observed $1/\gamma$ region (Jakobsson et al. 2004). The
$1/\gamma$ region has a transverse size of $\sim r/\gamma$ where $r$ is
the radial distance to the gamma-ray emitting region.  With $10^{14}
\le r \le 10^{16}$~cm the variations need only be somewhat smaller than
say $10^{12}$~cm.  Collisionless shocks mediated by the Weibel
instability have density and current structures with sizes on the order
of the electron skin depth. The typical transverse Weibel filament size
$\lambda_{ce} \sim \gamma^{1/2} c/\omega_{pe} \sim 3 \times
10^{11}/\omega_{pe}$,  where here the relevant value of the plasma
frequency in the observer's frame for the Weibel instability is
$\omega_{pe}/\gamma^{1/2}$. We note that the length of filaments will
be subject to length contraction but since the longitudinal plasma
frequency is $\omega_{pe}/\gamma^{3/2}$ the filament aspect ratio should
be preserved. The resultant size for any reasonable estimate of the
plasma frequency is many orders of magnitude smaller than the
asymmetric density variations implied by the light curves.  Thus, the
gamma-ray burst may be composed of emission from many different regions
but with variation from region to region on much larger scales than
those we have considered here. Since we found little difference between
no magnetic field and modest magnetic field (the Alfv\'en wave speed
was on the order of the thermal speed), we might expect our present
results to apply in the presence of magnetic fields with magnetic
energy in equipartition with the thermal energy in the downstream region.

The generation of magnetic fields both with and without an initial
magnetic field suggests that emission in GRB afterglows and Crab-like
pulsar winds could be either synchrotron or jitter emission (Medvedev
2000).  The size of filaments appear to be smaller than can produce
observable variations in intensity structure. However, this small size
can mean that the deflection angle, $\alpha \sim e B_{\perp}
\lambda_{\rm B}/\gamma m_e c^{2}$, of particles by Weibel filaments is
smaller than the radiation beaming angle, $\Delta\theta \sim 1/\gamma$
(Medvedev 2000).  Here $\lambda_{\rm B} \sim \lambda_{\rm ce}$, $e
B_{\perp}/m_e c <  \Omega_{\rm e}$, and the ratio $\delta \sim
\alpha/\Delta\theta < \Omega_{\rm e}/\omega_{\rm pe}$ will be less
than one when the cyclotron frequency is less than the plasma
frequency.  Thus, when ambient magnetic fields are moderate, i.e., the
cyclotron frequency is less than the plasma frequency and $\delta < 1$,
the emission may correspond to jitter rather than synchrotron
radiation.

Our simulation studies have provided a framework for the dynamics
of a relativistic shock generated within an electron-positron or an
electron-ion relativistic jet. The Lorentz factor $\gamma = 5$ set of
simulations is appropriate to internal shocks resulting from faster
material overtaking slower material in the reference frame of the
slower material, ambient medium in our simulations.  Here the ``shock''
Lorentz factor, $\gamma_{sh}$, can be related most simply to the Lorentz
factors of high, $\gamma_{h}$, and low, $\gamma_{\ell}$, speed material
with
$$
\gamma_{sh} = \epsilon \gamma_{h}^{2}
[1 - (1 - 1/\gamma_{h}^{2})^{1/2}(1 - 1/(\epsilon \gamma_{h})^{2})^{1/2}]
$$
where $1 \ge \epsilon \equiv \gamma_{\ell}/\gamma_{h} \ge
1/\gamma_{h}$.  Provided $\gamma_{h} >> 10$, $\gamma_{sh} \sim 5$ implies
$\epsilon \sim 1/10$.  For example, our present simulation set is relevant to
internal AGN jet shocks produced by $\gamma_{h} \sim 20$ material
overtaking $\gamma_{\ell} \sim 2$ material and to internal GRB shocks
produced by $\gamma_{h} \sim 300$ material overtaking $\gamma_{\ell}
\sim 30$ material.

The fundamental characteristics of relativistic shocks are essential
for a proper understanding of the prompt gamma-ray and afterglow
emission in gamma-ray bursts, and also to an understanding of the
particle reacceleration processes and emission from the shocked regions
in relativistic AGN jets.  Since the shock dynamics is complex and
subtle, more comprehensive studies are required to better understand
the acceleration of electrons, the generation of magnetic fields and
the associated emission. This further study will provide the insight
into basic relativistic collisionless shock characteristics needed to
provide a firm physical basis for modeling the emission from shocks in
relativistic flows.

\acknowledgments
K. Nishikawa is a NRC Senior Research Fellow at NASA Marshall Space
Flight Center. The author K.I.N. thanks useful discussions with
Christian Hededal. This research (K.N.) is partially supported by the
National Science Foundation awards ATM 9730230, ATM-9870072,
ATM-0100997, and INT-9981508. P. Hardee acknowledges partial support by
a National Space Science \& Technology (NSSTC/NASA) award.  The
simulations have been performed on ORIGIN 2000 and IBM p690 (Copper) at
the National Center for Supercomputing Applications (NCSA) which is
supported by the National Science Foundation.

\clearpage

\begin{figure}[ht]
\vspace*{2.0cm}
\plotone{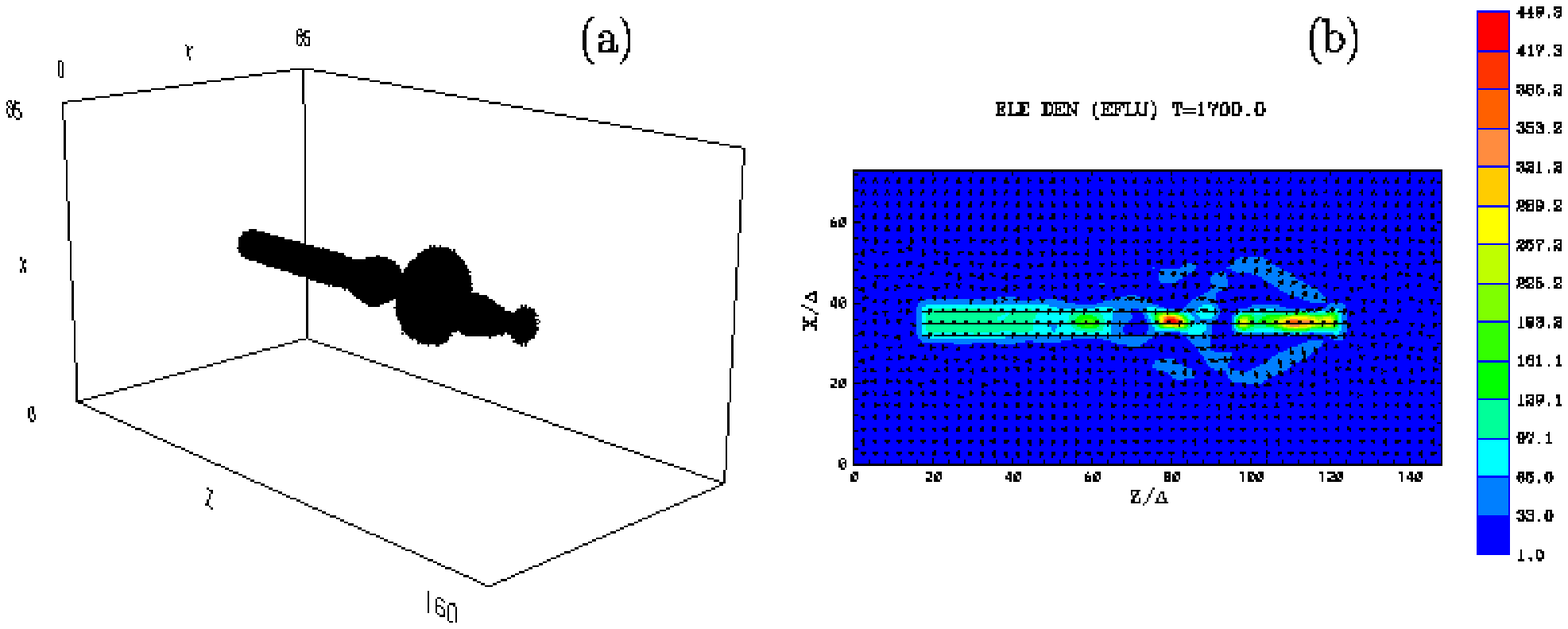}
\vspace*{-1.0cm}
\caption{Dynamics of a thin jet are indicated at $t = 22.1/\omega_{\rm pe}$
by (a) a jet electron image in the 3-dimensional simulation system, and
(b) the total electron density in the $x - z$ plane in the center of the jet
with the electron flux indicated by arrows and density indicated by color.}
\end{figure}

\clearpage

\begin{figure}[ht]
\vspace*{2.5cm}
\plotone{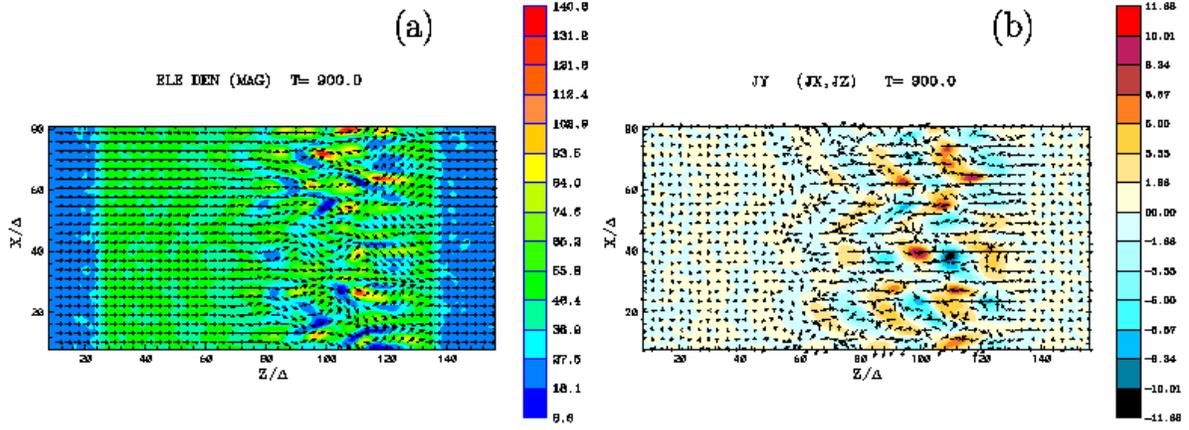}
\vspace*{-1.0cm}
\caption{2D images
in the $x - z$ plane at $y = 43\Delta$ for a flat jet injected into a 
magnetized ambient medium shown at $t = 23.4/\omega_{\rm pe}$.  In
(a) color indicates the electron density (peak: 140.6) with magnetic
fields represented by arrows and in (b) color indicates the $y$-component
of the current density ($J_{\rm y}$) (peak: 11.8) with $J_{\rm z},
J_{\rm x}$ indicated by the arrows.}
\end{figure}

\clearpage

\begin{figure}[ht]
\vspace*{2.0cm}
\plotone{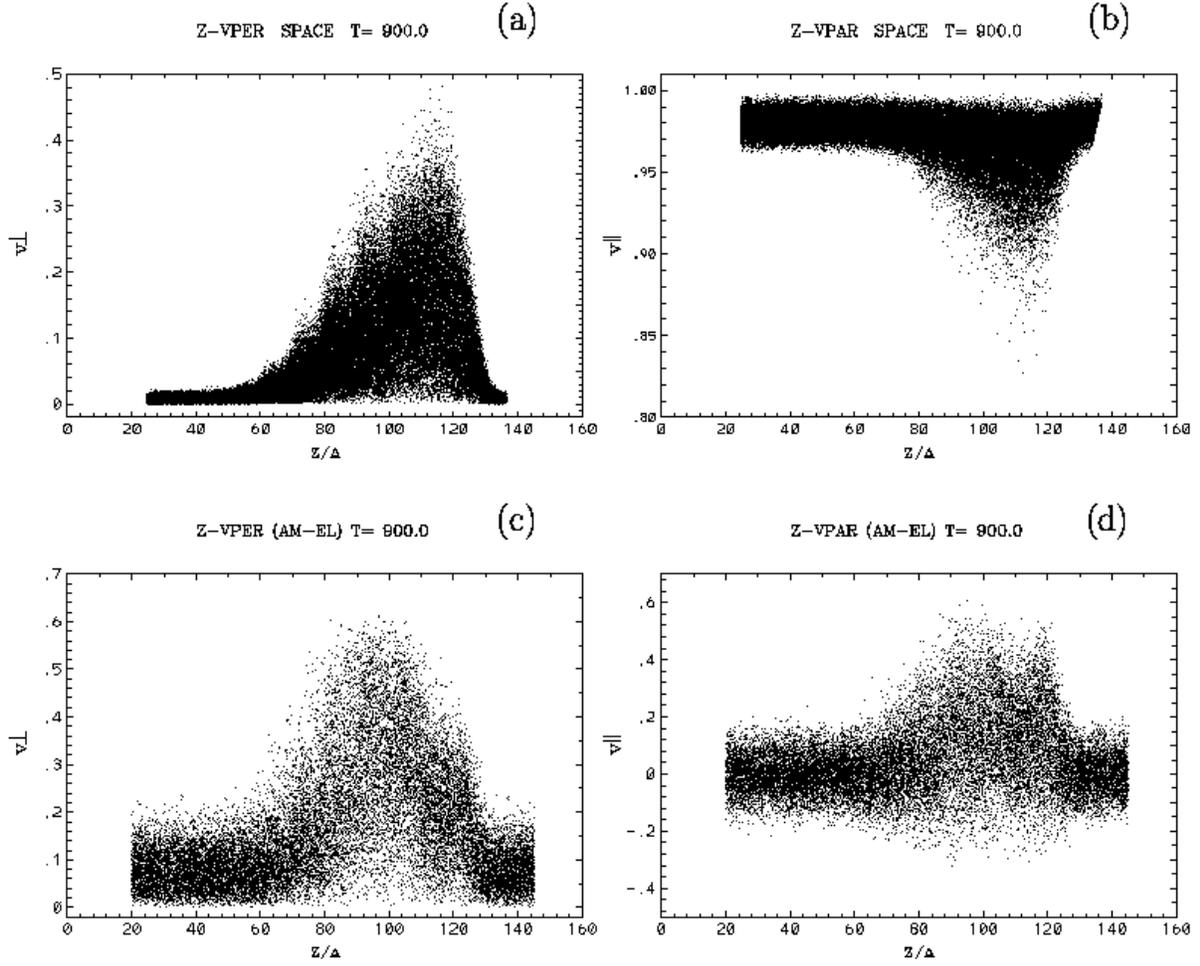}
\vspace*{-1.0cm}
\caption{The distribution of jet and ambient
electrons at $t = 23.4/\omega_{\rm pe}$ in (a) $z - v_{\perp}/c$
(jet), (b) $z - v_{\parallel}/c$ (jet), (c) $z - v_{\perp}/c$
(ambient), (d) $z - v_{\parallel}/c$ (ambient) phase space.
Roughly 20\% of the jet electrons and 0.1\% of the ambient
electrons ($20 < z/\Delta <145$) are randomly selected for these
plots.}
\end{figure}

\clearpage

\begin{figure}[ht]
\vspace*{-2.5cm}
\plotone{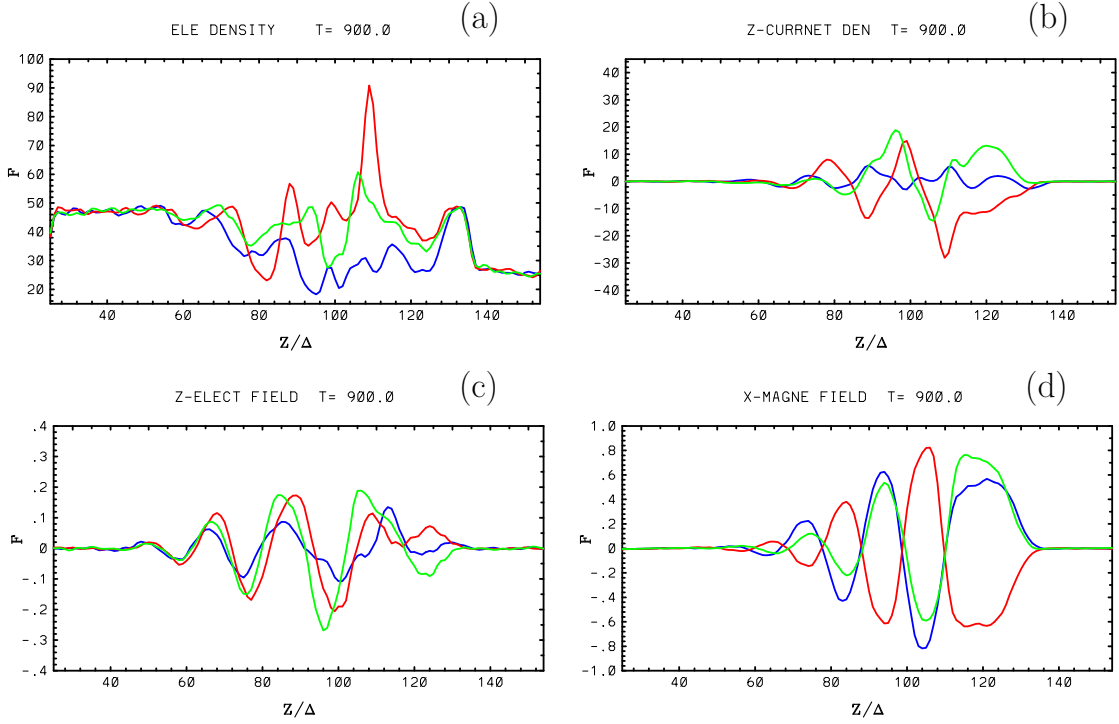}
\vspace*{-8.0cm}
\caption{One-dimensional cuts along the
$z$-direction ($25 \leq z/\Delta \leq 154$) of a flat
jet. Shown are (a) the electron
density, (b) the $z$-component of the current density, (c) the
$z$-component of the electric field, and (d) the $x$-component of
the magnetic field shown at $t = 23.4/\omega_{\rm pe}$. Cuts
are taken at $x/\Delta = 38$ and $y/\Delta = 38 (blue-dotted), 
43 (red-solid), 48 (green-dashed)$
and separated by about one electron skin depth.}
\end{figure}

\clearpage

\begin{figure}[ht]
\vspace*{1.0cm}
\plotone{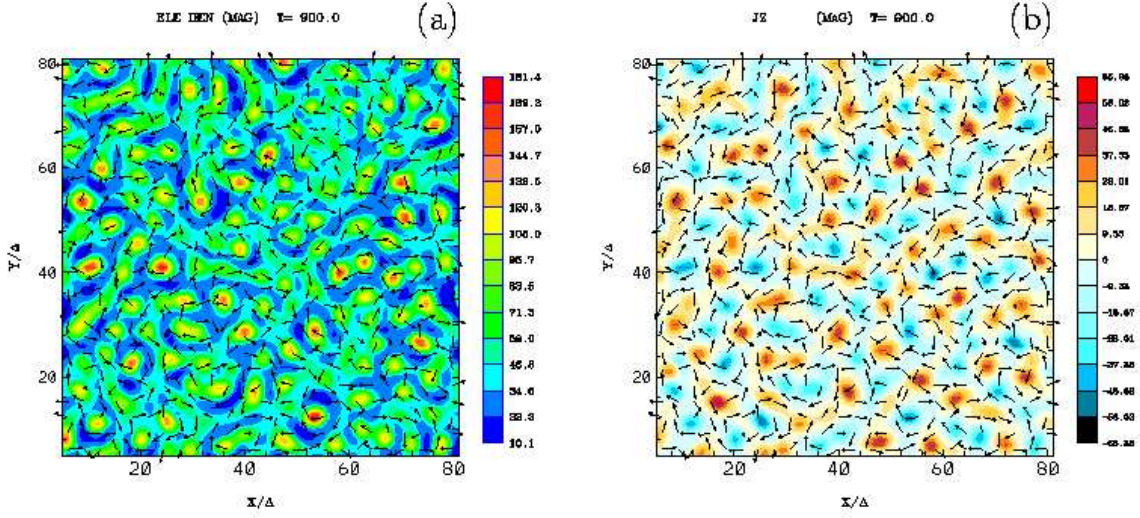}
\vspace*{-1.0cm}
\caption{The electron density (a) and z-component of the current density
(b) in the $x - y$ plane is plotted at $z/\Delta =$ 120 at 
$t = 23.4/\omega_{\rm pe}$. 
The arrows show the transverse magnetic fields $B_{\rm x,\, y}$.}
\end{figure}

\clearpage

\begin{figure}[ht]
\epsscale{1.00}
\plotone{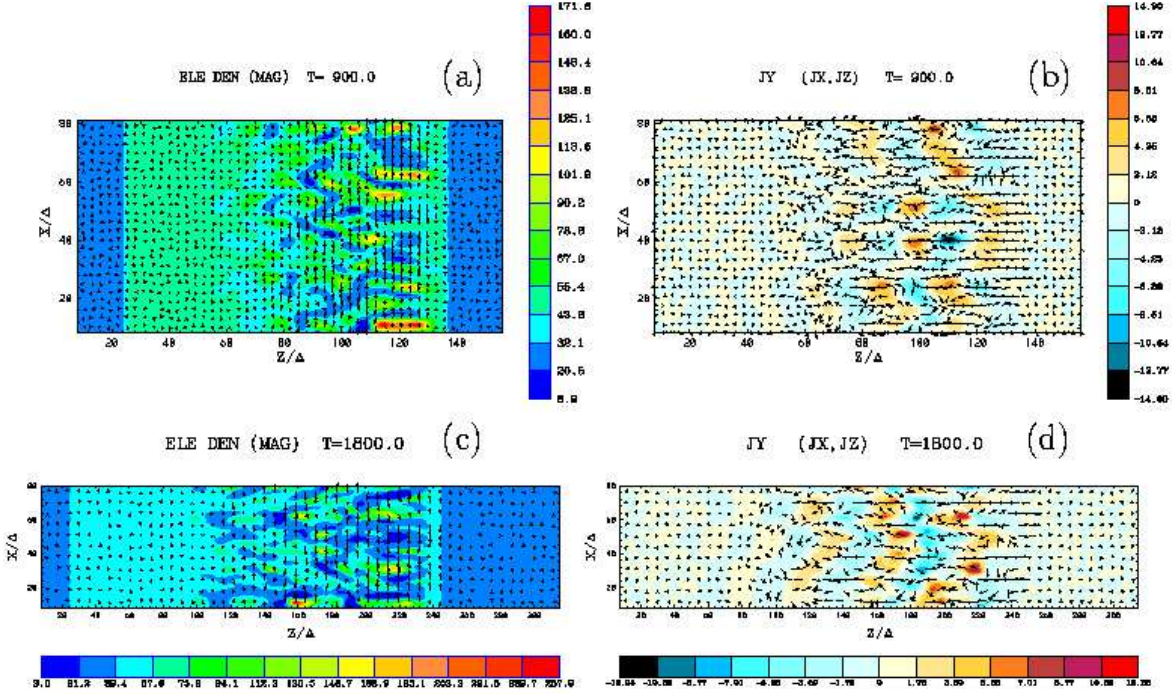}
\vspace*{-1.0cm}
\caption{2D images in the $x - z$ plane at $y = 43\Delta$ for a
flat jet injected into an unmagnetized ambient medium shown at
$t = 23.4/\omega_{\rm pe}$.  The effects of different electron
skin depth are shown ((a, b): 4.8$\Delta$;
(c, d): 9.6$\Delta$). Due to the longer skind depth, a longer 
jet was simulated in (c) and (d). In (a) and (c) color
indicates the electron density (peak: (a) 171.6, (c) 257.9) with 
magnetic fields indicated by arrows and in (b) and (d) color 
indicates the $y$-component of the current density $J_{\rm y}$ 
(peak: (b) 14.90,
(d) 12.28), with $J_{\rm z}, J_{\rm x}$ indicated by the arrows.
Images (a) and (b) are comparable to images in Fig.\ 2 
(magnetized) but the color
scales are different.}
\end{figure}

\clearpage

\begin{figure}[ht]
\vspace*{1.0cm}
\plotone{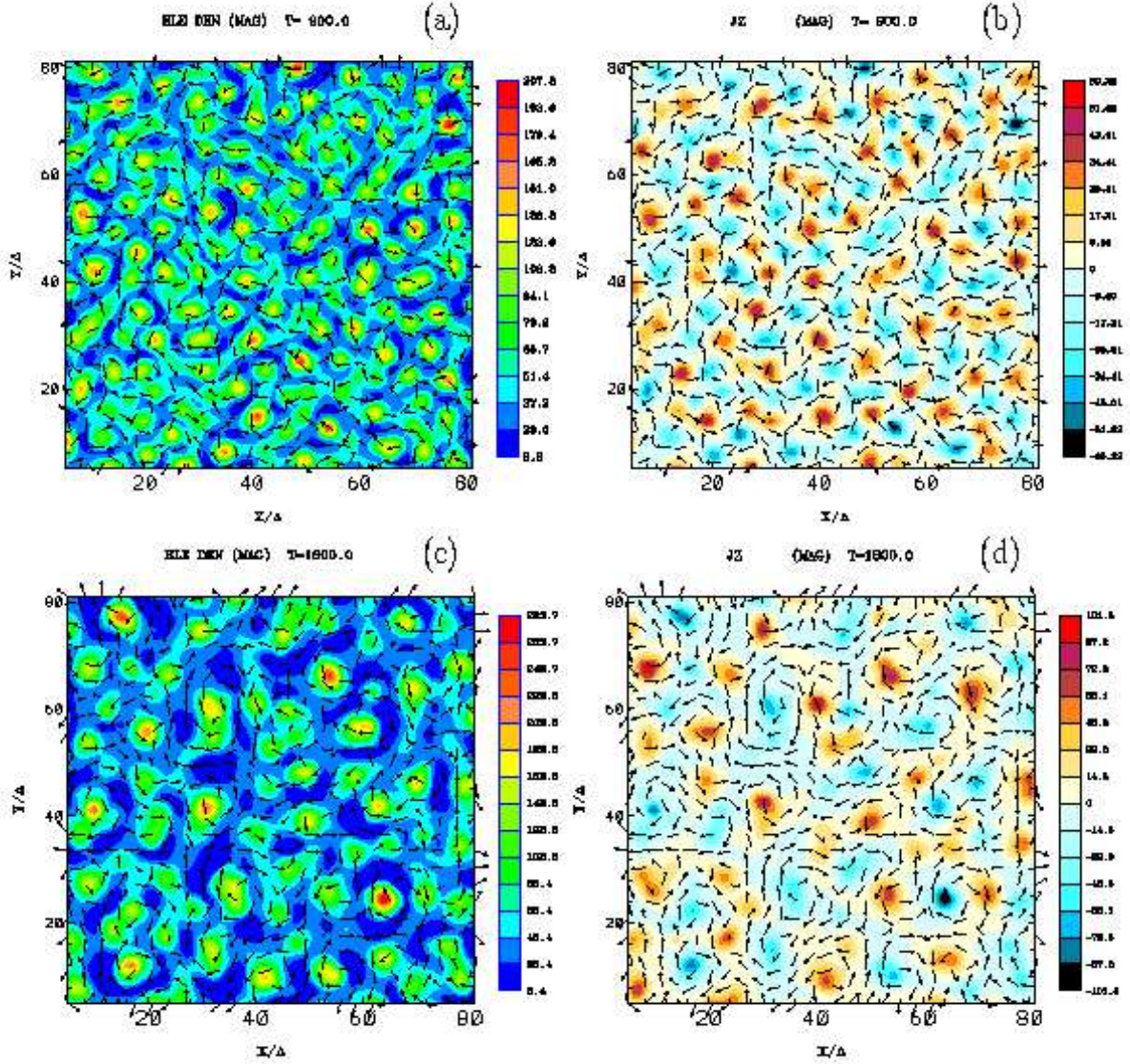}
\vspace*{-1.0cm}
\caption{2D images in the $x - y$ plane
for a flat jet injected into an unmagnetized ambient medium at
shown $t = 23.4/\omega_{\rm pe}$.  The effects of different electron skin
depth are shown and as in Fig.\ 6 ((a, b): 4.8$\Delta$;
(c, d): 9.6$\Delta$). The electron density (a, c)) and
$z$-component of the current density (b, d) are plotted
at $z = 120\Delta$ (a, b) and $z = 215\Delta$ (c, d). The arrows
show the transverse magnetic fields $B_{\rm x,\, y}$ which are generated
by $J_{\rm z}$ (in particular Fig. 7d).}
\end{figure}

\clearpage

\begin{figure}[ht]
\vspace*{-2.5cm}
\plotone{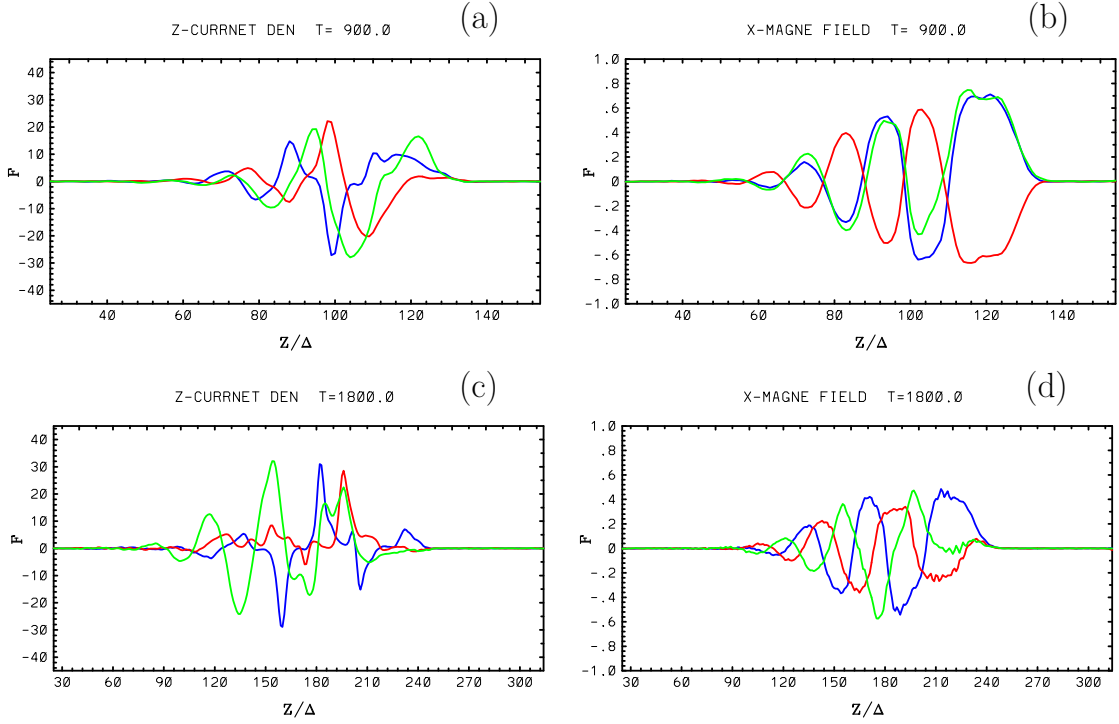}
\vspace*{-8.0cm}
\caption{One-dimensional cuts in the $z$-direction of the current density 
($z$-direction) (a, c), and the magnetic field ($x$-component) (b, d) 
shown at $t = 23.4/\omega_{\rm pe}$.  The effects of different electron skin
depth are shown and as in Figs.\ 6 \& 7 ((a, b): 4.8$\Delta$;
(c, d): 9.6$\Delta$). Cuts are taken at $x/\Delta = 38$
and (a, b) $y/\Delta = 38 (blue-dotted), 43 (red-solid), 48 
(green-dashed)$ or (c, d)
$y/\Delta = 33 (blue-dotted), 43 (red-solid), 53 (green-dashed)$,
and are separated by about one electron skin depth.}
\end{figure}

\end{document}